\documentclass[11pt,a4paper,eps]{article}
\usepackage{graphicx,amssymb}
\usepackage{amsmath}
\usepackage{graphicx}
\usepackage{amsfonts}
\usepackage{amssymb}
\begin{document}



\begin{center}
{\Large\bf Black-hole thermodynamics
with modified dispersion relations and generalized uncertainty principles}
\end{center}

\vskip 0.5 cm

\begin{center}

{\bf Giovanni AMELINO-CAMELIA}$^a$, {\bf Michele ARZANO}$^b$,\\
 {\bf Yi LING}$^{c}$ and {\bf Gianluca MANDANICI}$^a$\\

\end{center}

\begin{center}
{\small $^a${\it Dip.~Fisica Univ.~Roma ``La Sapienza'' and
Sez.~Roma1 INFN,}}\\{\small $~${\it Piazzale Moro 2, Roma, Italy}}\\
{\small $^b${\it Institute of Field Physics, Dept Physics and
Astronomy}}\\{\small $~${\it University of North Carolina, Chapel
Hill, NC 27599,
USA}}\\
{\small $^c${\it Center for Gravity and Relativity, Department of
Physics,}}\\{\small $~${\it Nanchang University, Nanchang 330047, China}}
\end{center}

\vskip 0.5 cm


\begin{center}

{\bf SUMMARY}

\end{center}

\begin{quotation}

\leftskip=0.6in \rightskip=0.6in

{\small In several approaches to the quantum-gravity problem
evidence has emerged of the validity of a ``GUP" (a Generalized
position-momentum Uncertainty Principle) and/or a ``MDR"
(a modification of the energy-momentum dispersion
relation), but very little is known about the implications of GUPs
and MDRs for black-hole thermodynamics, another key topic for
quantum-gravity research. We investigate an apparent link, already
suggested in an earlier exploratory study involving two of us,
between the possibility of a GUP and/or a MDR and the possibility
of a log term in the area-entropy black-hole formula. We then
obtain, from that same perspective, a modified relation between the
mass of a black hole and its temperature, and we examine the
validity of the ``Generalized Second Law of black-hole
thermodynamics" in theories with a GUP and/or a MDR. After an
analysis of GUP- and MDR-modifications of the black-body radiation
spectrum, we conclude the study with a description of the
black-hole evaporation process.}
\end{quotation}

\baselineskip 12pt plus .5pt minus .5pt

\section{Introduction}
Various arguments suggest that the description of black holes
should be an important aspect of a quantum-gravity theory, and
that some key operatively-meaningful (not merely formal)
differences between alternative theories should emerge as we
establish, within each approach, how the singularity, the
evaporation, the ``thermodynamics'', and the ``information
paradox'' are handled. A similar role could be played, as stressed
in the recent literature~\cite{grbgac,3perspe,smoliREV,daviREV},
by the analysis of the
energy-momentum dispersion relation and the position-momentum
uncertainty principle. Different approaches to the quantum-gravity
problem lead to different expectations for what concerns the
possibility of a MDR (a modified energy-momentum dispersion
relation) and the possibility of a GUP (a generalized
position-momentum uncertainty principle). In particular, in the
study of  Loop Quantum Gravity and of models based on
noncommutative geometry there has been strong
interest~\cite{lqgDispRel1,lqgDispRel2,lukieAnnPhys}
in some candidate modifications of the
energy-momentum dispersion relation. Generalized uncertainty
principles\cite{Maggiore:1993rv,Garay:1994en}
have been considered primarily in the literature on
String Theory~\cite{venegross} and on models based on
noncommutative geometry~\cite{dopl1994}.
The form of the energy-momentum dispersion relation and of the
position-momentum uncertainty relation can therefore be used to
characterize alternative approaches to the quantum-gravity
problem.

Two of us were recently involved~\cite{aap} in research exploring
a possible link between the predictions that a quantum-gravity
theory makes for black-hole thermodynamics and the predictions
that the same theory makes for the energy-momentum dispersion
relation and the position-momentum uncertainty relation. By
establishing the nature of such a link one would, in our opinion,
obtain a valuable characterization of the type of internal logical
consistency that various aspects of a quantum-gravity theory
should satisfy. The study reported in Ref.~\cite{aap} was not the
first to explore the possible role of MDRs and/or GUPs for black
holes, as we discuss in greater detail here in Section 9 where we
comment of some relevant
references~\cite{Blaut:2001fy,Adler:2001vs,calib2,Chang:2001bm,Yepez:2004sa,am1},
but all of these studies, including Ref.~\cite{aap}, focused on one or
another aspect of the possible interplay between MDR/GUP results
and black holes, without attempting to obtain a wider picture. We
here work in the spirit of Ref.~\cite{aap}, but we attempt to give
the first elements of a general analysis of some key
characteristics of black-hole physics, as affected by some
scenarios for a MDR or a GUP.

Sections 2, 3 and 4, set the stage, by reviewing some results in the MDR/GUP
literature and revisiting the point already made in Ref.~\cite{aap},
which concerns an apparent link between
the log-area terms in the entropy-area relation
for black holes and certain formulations of the MDR and the GUP.
In Section 5 we explore, still within the working assumptions adopted in
Ref.~\cite{aap}, the implications of a MDR and/or
a GUP for the Bekenstein entropy bound and
for the Generalized Second Law of thermodynamics. We find that the implications
are significant and we conjecture that they should also not be negligible in the
analysis of other entropy-bound proposals.
Section 6 considers a role for MDR/GUP modifications in the analysis
of the black-body radiation spectrum, and again exposes some significant changes
with respect to the standard picture, including the possibility that the
characteristic frequency of black-body radiation at given temperature $T$
might have a dependence on $T$ such that in the infinite-temperature limit
the characteristic frequency would take a finite (Planckian) value.
It is then perhaps not surprising that in the analysis of the black-hole evaporation
process, discussed in Section 7, we also find some characteristic MDR/GUP-induced
new features, such as the possibility that the energy flux emitted by the black hole
might diverge when the black-hole mass reaches a certain finite (Planckian) value.
In Section 8, we comment on one key aspect which might deserve further consideration:
for these theories with MDRs and/or GUPs there has been some speculation that
the speed of massless particles might be different from the familiar
speed-of-light scale value of $c$. In Sections 1-7 we assume throughout
that $c$ still is the speed
of massless particles, but in Section 8 we establish how the analysis
of black-body radiation would be changed if one implemented some alternatives considered
in the literature.
In Section~9 we compare our analysis with other studies which have considered the
implications of a MDR or a GUP for some aspects of black-hole physics.
Section 10 concludes the paper with some remarks on the outlook of this
research programme.

\section{MDRs and GUPs in Quantum Gravity and implications for a Planck-scale
particle-localization limit}

\subsection{MDRs and GUPs in Quantum Gravity}
In the study of the Quantum-Gravity problem the emergence of modified
energy-momentum relations and/or generalized position-momentum uncertainty
principles, although of course not guaranteed, can be motivated
on general grounds, and also finds support in the direct
analysis of certain Quantum-Gravity scenarios.

The hypothesis of modified energy-momentum dispersion relations
is understandably popular among those adopting a ``spacetime foam" intuition
in the study of the quantum-gravity problem, especially when an analogy between
spacetime foam and some more familiar forms of medium (such as certain crystal
structures of interest in condensed-matter studies) is proposed.
It is then expected that wave dispersion ``in vacuo" (in the spacetime foam)
might resemble wave dispersion in other media.
A modified dispersion relation can also be favoured by the
expectation, shared by many researchers of the field, that the Planck length
might fundamentally set the minimum allowed value for wavelengths.
A nonlinear relationship between energy and (space-) momentum can be easily
adjusted in such a way that in the infinite-energy limit the momentum
saturates to the Planck-scale value (and wavelength saturates to the Planck-length
value). This possibility has become more attractive with
the recent realization~\cite{gacdsr,jurekdsr,leedsr} that a modified energy-momentum
dispersion relation can also be introduced as an observer-independent
law\footnote{But this usually requires introducing a nonlinear deformation
of the action of Lorentz boosts.}, in which case the Planckian minimum-wavelength
hypothesis can be introduced as a physical law valid in every frame.
The analysis of some quantum-gravity scenarios, even in cases in which the emergence
of modified energy-momentum relations was not intended in the original
setup of the framework, has shown some explicit mechanisms for the emergence
of modified dispersion relations. This is particularly true of some
approaches based on noncommutative geometry~\cite{lukieAnnPhys,jurekdsr}
and within the Loop-Quantum-Gravity
approach~\cite{lqgDispRel1,lqgDispRel2}.
In most cases one is led to consider a dispersion relation of the
type\footnote{We denote with $m$, as conventional, the rest energy
of the particle. The mass parameter $\mu$ on the right-hand side
is directly related to the rest energy, but $\mu \neq m$ if the $\alpha_i$
do not all vanish. For example, if $\alpha_1 \neq 0$ but  $\alpha_i = 0$
for every $i \geq 2$ one of course obtains $\mu^2 =  m^2 + \alpha_1 L_p m^3$.
This needed to be clarified since it is relevant for more general analyses
of MDRs, but in our study we are always concerned with particles which
are either massless or anyway are analyzed at energies such that the mass
can be neglected, and therefore both $\mu$ and $m$ will never actually enter
our key formulas.}
\begin{equation}
 \vec{p}^2 = f(E,m;L_p) \simeq E^2 - \mu^2+ \alpha_1 L_p E^3
+ \alpha_2 L_p^2 E^4+ O\left(L_p^3 E^5 \right) ~, \label{disprelONE}
\end{equation}
where $f$ is the function that gives the exact dispersion relation, and on the
right-hand side we just assumed the applicability of a Taylor-series
expansion for $E \ll 1/L_p$. The coefficients $\alpha_i$ can take different
values in different Quantum-Gravity proposals.

The fact that these Planck-scale-deformed dispersion relations
may have observably large consequences in some (however rare) physical contexts
has led to interest in this research also from the perspective of
phenomenology~\cite{grbgac,kifu,ita,gactp,jaco}.

The situation concerning the possibility
of a generalized position-momentum uncertainty principle is rather similar.
On general grounds it can be motivated by the intuition\cite{Maggiore:1993rv,Garay:1994en}
that the solution
of the quantum-gravity problem might require the introduction
of an absolute
Planckian limit on the size of the collision region, applicable
to high-energy microscopic collision processes. For example, a GUP of the form
\begin{equation}
\delta x \ge \frac{1}{\delta p} +\alpha L_p^2 \delta p + O(L_p^3 \delta p^2)~,
\label{GUP}
\end{equation}
which has been derived within the String Theory approach to the quantum-gravity
problem~\cite{venegross},
is such that at small $\delta p$ one finds the standard dependence
of $\delta x$ on $\delta p$ ($\delta x$ gets smaller as $\delta p$ increases)
but for large $\delta p$ the Planckian correction term becomes significant
and keeps $\delta x \ge L_p$. Within String Theory the coefficient $\alpha$
should take a value of roughly the ratio between the square of the string length and
the square of the Planck length, but this of course might work out differently
in other Quantum-Gravity proposals.

While in the parametrization of (\ref{disprelONE}) we included a possible
correction term suppressed only by one power of the Planck length,
in (\ref{GUP}) such a linear-in-$L_p$ is assumed not to be present.
This reflects the status of the presently-available literature:
for the MDR a large number of alternative formulations, including some
with the linear-in-$L_p$ term, are being considered, as they find support
in different approaches to the quantum-gravity problem (and different
preliminary results adopting alternative approximation schemes within
a given approach), whereas all the discussions of a GUP assume that
the leading-order
correction should be proportional to the square of $L_p$.

\subsection{MDR and a Planck-scale particle-localization limit}
The analysis reported in Ref.~\cite{aap} exposed a previously unnoticed common
feature of MDR and GUP scenarios. This has to do with a Planck-scale limit
on the localization of a particle, and an associated modification of the Bekenstein
argument for a area-entropy black-hole relation.

Arguably the closest starting point for the construction of the correct
Quantum Gravity should be Quantum Field Theory, and within Quantum Field Theory
the most striking quantum effect concerns an absolute limit on the localization
of a particle of energy $E$, codified in the
relation $E\ge\frac{1}{\delta x}$.
While in nonrelativistic quantum mechanics a particle of any energy can always
be sharply localized (at the price of renouncing to all information
on the conjugate momentum), within Quantum Field Theory only in the infinite-energy
limit a particle can be sharply localized.
And among those studying the quantum-gravity problem one frequently
encounters the intuition that at the Quantum-Gravity level the idealization
of sharp localization should disappear completely.

In the spirit of Ref.~\cite{aap} one can attempt to codify this quantum-gravity
intuition in a relation of the type
\begin{equation}
E\ge\frac{1}{\delta x} \left(1- \Delta(L_p,\delta x) \right)
\label{loclim}
\end{equation}
where $\Delta$ is some function of $L_p$ and $\delta x$, perhaps
such that $E \rightarrow \infty$ already at some finite value of $\delta x$
(so that the idealization $\delta x \rightarrow 0$ is excluded).
And it was observed in Ref.~\cite{aap} that both the idea of a MDR and the idea
of a GUP would support a formula of the type (\ref{loclim}), with nonzero $\Delta$.

Let us briefly review this analysis reported in Ref.~\cite{aap}, starting with the
case of a MDR of the type (\ref{disprelONE}).
We can follow the familiar derivation~\cite{landau}
of the relation $E\ge\frac{1}{\delta x}$, substituting, where necessary, the standard
special-relativistic dispersion relation with
its Planck-scale modified version. It is convenient to start
by focusing on the case of a particle of
mass $M$ at rest, whose
position is being measured by a procedure
involving a collision with a photon of
energy $E_\gamma$ and momentum $p_\gamma$.
According to Heisenberg's uncertainty principle,
in order to measure the
particle position with precision $\delta x$ one should use a photon with
momentum uncertainty $\delta p_\gamma\ge\frac{1}{\delta x}$.
Following the standard argument~\cite{landau}, one takes
this $\delta p_\gamma\ge\frac{1}{\delta x}$ relation and
converts it into the relation $\delta E_\gamma\ge\frac{1}{\delta x}$
using the special relativistic dispersion
relation. Finally $\delta E_\gamma\ge\frac{1}{\delta x}$ is
converted into the relation  $M\ge\frac{1}{\delta x}$
because the measurement procedure requires  $\delta E \le M$,
in order to ensure that the relevant energy
uncertainties are not large enough to allow the production of
additional copies of the particle whose position is being measured.

If indeed our Quantum-Gravity scenario hosts
a Planck-scale modification of the dispersion relation of the form
(\ref{disprelONE}) then clearly the relation between $\delta p_\gamma$
and $\delta E_\gamma$ should be
 re-written as follows
\begin{equation}
\delta p_\gamma \simeq  \left(1+ \alpha_1 L_p E+3\left(\frac{\alpha_2}{2}-\frac{\alpha_1^2}{8}\right)
L_p^2 E^2\right)\delta E_\gamma
\end{equation}
which then leads to the requirement
\begin{equation}
M \ge \frac{1}{\delta x} - \alpha_1 \frac{L_p}{(\delta x)^2}
+\left(\frac{11}{8}  \alpha_1^2-\frac{3}{2} \alpha_2 \right) \frac{L_p^2}{(\delta x)^3} +
O\left(\frac{L_p^3}{(\delta x)^4}\right)\, .
\end{equation}

These results strictly apply to the measurement
of the position of a particle at rest, but they can be
straightforwardly generalized~\cite{landau}
(simply using a boost) to the case of the measurement of the position
of a particle of energy $E$.
For the standard case this leads to the $E \ge 1/ \delta x$ relation
while in presence of an MDR one easily finds
\begin{equation}
E \ge \frac{1}{\delta x} - \alpha_1 \frac{L_p}{(\delta x)^2}
+\left(\frac{11}{8}  \alpha_1^2-\frac{3}{2} \alpha_2 \right) \frac{L_p^2}{(\delta x)^3} +
O\left(\frac{L_p^3}{(\delta x)^4}\right)\, .
\label{deform}
\end{equation}

\subsection{GUP and a Planck-scale particle-localization limit}
While the connection between a MDR and a Planck-scale particle-localization limit
is somewhat less obvious (and in fact we found no mention of it in the literature
previous to Ref.~\cite{aap}),
it is not at all surprising that the GUP would give rise to such
a particle-localization limit.
In fact, as mentioned, the GUP is primarily viewed as a way to introduce a
Planckian limit on the size of the collision region, applicable
to high-energy microscopic collision processes, and a limitation
on the size of collision regions would naturally be expected to
lead to a particle-localization limit.
Indeed, as the careful reader can easily verify,
from the GUP one obtains (following again straightforwardly
the familiar line of analysis discussed in Ref.~\cite{landau})
a modification of the relation $E  \ge 1/\delta x$.
The modification is of the type $E  \ge 1/\delta x + \Delta$,
with $\Delta$ of order $\alpha L_p^2/\delta x^3$,
and originates from the fact that
according to the GUP, (\ref{GUP}), one
obtains  $\delta p_\gamma \ge 1/\delta x +\lambda_s^2/\delta x^3$
(instead of the original $\delta p_\gamma \ge 1/\delta x$).
Using the standard special-relativistic dispersion relation for
a photon $p_\gamma=E_\gamma$ the condition on the momentum
uncertainty translates in a condition on the energy
uncertainty $\delta E_\gamma\ge\frac{1}{\delta x}\left(1
+\alpha\frac{L_p^2}{\delta x^2}\right)$, and ultimately this leads to
\begin{equation}
E \ge \frac{1}{\delta x} + \alpha \frac{L_p^2}{(\delta x)^3}
+ O\left(\frac{L_p^3}{(\delta x)^4}\right)\, .
\label{deformGUP}
\end{equation}

\section{MDR and black hole entropy}
In this section we revisit the argument already proposed in Ref.~\cite{aap},
suggesting that a Planck-scale modification of the particle-localization
limit, of the type (\ref{deform}) or (\ref{deformGUP}),
can be used to motivate corrections to
the $S =A/(4 L_p^2)$ area-entropy relation for black holes.
We focus here on the case of a MDR, but since the key ingredient
is the Planck-scale particle-localization
limit, one should expect that, as we confirm explicitly in the next section,
the same line of analysis is applicable also to the case in which
one takes as starting point a GUP.
Since the literature on MDRs is composed both of papers using arguments
that are only sufficient to specify the first terms in a series expansion of
the MDR at energies below the Planck scale, and some analyses proposing a complete
all-order formula for the MDR, we find appropriate to consider these possibilities
separately.
The power-series analysis will already show that the implications of a MDR
can be significant. But the power-series analysis can only be reliably used
at energies safely below the Planck scale.
In considering some examples of all-order MDR proposals we will
also develop some intuition for the type of implications that a MDR
could have for black-hole physics at Planckian energy scales.

Our argument connecting a MDR (a particle-localization
limit)
and some modifications of the area-entropy relation for black holes
is formulated in a scheme of analysis first introduced by Bekenstein~\cite{bek},
which is actually one of the classic arguments for the description
of the entropy-area relation.
In order to render our presentation self-contained we open this
section by describing this classic Bekenstein argument,
but we just sketch out the Bekenstein derivation
since we expect most readers to be already familiar with it.

\subsection{The original Bekenstein argument (with unmodified dispersion relation
and unmodified uncertainty principle)}
The argument presented by Bekenstein in Ref.~\cite{bek}
uses very simple ingredients to suggest that
the entropy of a black hole should be proportional
to its (horizon-surface) area.
The starting point is the general-relativity result~\cite{christo}
establishing
that the minimum increase of area when the black hole absorbs a classical
particle of energy $E$ and size $s$ is $\Delta A \simeq 8 \pi L_p^2 E s$
(in ``natural units" with $\hbar=c=1$).
In order to describe the absorption
of a quantum particle one must describe the size of the
particle in terms of the uncertainty in its
position~\cite{bek,hod}, $s \sim \delta x$,
and take into account a ``calibration
factor"~\cite{chenproc,calib3,aap} $(\ln 2)/2 \pi$
that connects the $\Delta A \ge 8 \pi L_p^2 E s$
classical-particle result with the quantum-particle
estimate $\Delta A \ge 4 (\ln 2) L_p^2 E \delta x$.
Bekenstein then enforces the requirement that
a particle with position uncertainty  $\delta x$
should at least~\cite{landau} have energy $E \sim  1/ \delta x$,
which leads to $\Delta A \ge 4 (\ln 2) L_p^2$, and
assumes that the entropy depends only on the area of the
black hole. Also using the fact that
the minimum increase of entropy should
be $\ln 2$, independently of the value of the area,
one then concludes that
\begin{equation}
\frac{dS}{dA} \simeq \frac{min (\Delta S)}{min (\Delta
A)} \simeq  \frac{\ln 2}{4 (\ln 2)  L_p^2}
~. \label{minDaADDED}
\end{equation}
From this it follows that
(up to an irrelevant constant contribution to entropy):
\begin{equation}
S \simeq \frac{A}{4 L_p^2}  ~. \label{entropy1ADDED}
\end{equation}

\subsection{MDR and black hole entropy in leading order}
The Bekenstein argument implicitly assumes (through the $E \ge 1/\delta x$
relation) that the energy-momentum dispersion relation and the
position-momentum uncertainty principle take the standard form.
Let us now reformulate the argument, still assuming a standard
form for the position-momentum uncertainty principle, but introducing
a MDR of the type (\ref{disprelONE}).
As in the original Bekenstein
argument~\cite{bek}, we take as starting
point the general-relativity result which
establishes that the area of a
black hole changes according to $\Delta A \ge 8 \pi L_p^2 E s $
when a classical particle of energy $E$ and size
$s$ is absorbed. And again we describe the size of the
particle in terms of the uncertainty in its position
as done in the previous subsection,
obtaining $\Delta A \ge 4 (\ln 2) L_p^2 E \delta x$.
Whereas in the original Bekenstein argument
one then enforces the relation $E \ge 1/ \delta x$ (and
this leads to $\Delta A \ge 4 (\ln 2) L_p^2$), we
must take into account the MDR-induced Planck-length modification
in (\ref{deform}), obtaining
\begin{eqnarray}
\Delta A & \ge & 4 (\ln 2) \! \left[L_p^2  - \!  \frac{\alpha_1 L_p^3}{\delta x}
- \! \frac{\left(\frac{3}{2}\alpha_2-\frac{11}{8}\alpha_1^2\right)
L_p^4}{(\delta x)^2} \right]\\ \nonumber
 & \! \simeq&
 4 (\ln 2) \! \left[ L_p^2  - \! \frac{\alpha_1 L_p^3}{R_S}
- \! \frac{ \left(\frac{3}{2}\alpha_2-\frac{11}{8}\alpha_1^2\right) L_p^4}{(R_S)^2}  \right]\\ \nonumber
 & \! \simeq &
4 (\ln 2) \! \left[ L_p^2 - \! \frac{\alpha_1
2 \sqrt{\pi} L_p^3}{\sqrt{A}} - \! \frac{\left(\frac{3}{2}\alpha_2-\frac{11}{8}\alpha_1^2\right)
4 \pi L_p^4}{A}  \right]
~,
\label{minDagac}
\end{eqnarray}
where we also used the fact that in falling in the black hole
the particle acquires~\cite{calib2,smaller,bigger}
position uncertainty $\delta x \sim R_S$, where $R_S$ is the
Schwarzschild radius (and
of course $A = 4 \pi R_S^2$).
From (\ref{minDagac}) we derive an area-entropy relation
assuming that the entropy of the black hole depends only on its area and
that the minimum increase of entropy should be,
independently of the value of the area, $\ln 2$:
\begin{equation}
\frac{dS}{dA} \simeq \frac{min (\Delta S)}{min (\Delta
A)} \simeq  \frac{\ln 2}{4 (\ln 2)  L_p^2 \left[
 1 - \! \frac{\alpha_1
2 \sqrt{\pi} L_p}{\sqrt{A}} - \! \frac{\left(\frac{3}{2}\alpha_2-\frac{11}{8}\alpha_1^2\right)
4 \pi L_p^2}{A}
\right]} \simeq
\left(\frac{1}{4 L_p^2} +  \frac{\alpha_1 \sqrt{\pi}}{2 L_p \sqrt{A}}
+  \frac{\left(\frac{3}{2}\alpha_2-\frac{11}{8}\alpha_1^2\right)  \pi}{A}\right)
~, \label{minDa}
\end{equation}
which gives (up to an irrelevant constant contribution to entropy)
\begin{equation}
S \simeq \frac{A}{4 L_p^2} + \alpha_1  \sqrt{\pi} \frac{\sqrt{A}}{L_p}
+ \left(\frac{3}{2}\alpha_2-\frac{11}{8}\alpha_1^2\right)  \pi \ln
\frac{A}{L_p^2} ~. \label{entropy1}
\end{equation}

This result of course reproduces the famous linear formula
if all coefficients $\alpha_i$ vanish.
If the cubic term $\alpha_1 E^3$ is present in the
energy-momentum dispersion relation then the
leading correction goes like $\sqrt{A}$, whereas if the first
nonzero coefficient in the dispersion relation
expansion is $\alpha_2$ the leading correction term goes
like $\log A$.
Our ``improved Bekenstein argument" therefore provides a possible
link between the form of the MDR (and of the GUP, as we stress later)
and the all-order form of the entropy-area relation for black holes.
For example, if within a given quantum-gravity approach
one can find a general argument suggesting that there are no $\sqrt{A}$
terms in the entropy-area relation, then one can use our
improved Bekenstein argument
to deduce that within that given quantum-gravity approach
one should not find terms of the type $\alpha_1 E^3$ in the
energy-momentum dispersion relation.

Over the last few years
both in String Theory and in Loop Quantum Gravity
some techniques for the direct analysis of the entropy of black holes,
using their quantum properties, have been developed,
and these techniques are now able \cite{stringbek,lqgbek}
to go even beyond the
entropy-area-proportionality contribution:
they establish that the leading correction should
be of log-area type, so that one expects (for $A \gg L_p^2$)
an entropy-area relation for black holes of the type
\begin{equation}
S = \frac{A}{4 L_p^2}
+ \rho \ln \frac{A}{L_p^2} + O\left(\frac{L_p^2}{A}\right)
~.
\label{linPLUSlog}
\end{equation}
where $\rho$ is a coefficient which might take different
value~\cite{stringbek,lqgbek,nonUniv,aap}
in String Theory and in Loop Quantum Gravity.
The status of the energy-momentum dispersion relation within these
theories is not completely settled (it is much debated particularly
in the Loop-Quantum-Gravity literature), but
on the basis of our improved
Bekenstein argument we can conclude that both
String Theory and Loop Quantum Gravity cannot allow
terms of the type $\alpha_1 E^3$ in the
energy-momentum dispersion relation.
And, if their log-area corrections
to the entropy-area relation are to be trusted,
we expect that both String Theory and Loop Quantum Gravity
should either predict a $\alpha_2 E^4$ correction to the
dispersion relation or (see later) they should host a corresponding
modification of the position-momentum uncertainty principle.

We can also use (\ref{entropy1}) to obtain,
using the first law of black hole thermodynamics $dS=\frac{dM}{T}$,
a Planck-scale-corrected relation between black-hole temperature
and mass:
\begin{equation}
T_{BH}^{MDR}\simeq\frac{E_p^2}{8\pi M}\left(1
-\alpha_1\frac{E_p}{2\sqrt{2}M}-\left(\frac{15}{32}\alpha_1^2-\frac{3}{8}\alpha_2\right)\frac{E_p^2}{M^2}\right),
\end{equation}
where we also used the familiar
relation between black hole area and mass $A=16\pi M^2$.

\subsection{Some all-order results for MDR modifications of black-hole entropy}
In the previous subsection we were establishing a possible relation between
MDR and log corrections to the entropy-area relation. Since the log-area
term is a leading-order term it was appropriate to
work within a power-series expansion of the MDR.
Moreover, the mentioned results from quantum-gravity research (primarily
from Loop Quantum Gravity and approaches based on noncommutative geometry)
that provide motivation for a Planck-scale modification of the dispersion
relation in most cases are obtained within analyses that only have
access to the first terms in a power-series expansion of the dispersion
relation. Still for some aspects of our analysis it will be useful
to contemplate some illustrative examples of all-order dispersion relations,
especially when we try to figure out what could be some examples of
implications of a MDR for the behaviour of black holes of Planck-length size.

The careful reader can easily verify that once a given energy-momentum
dispersion relation $E=f_{disp}(p)$ is adopted the steps of the
calculation reported in the preceding subsection can be followed
rather straightforwardly, obtaining
\begin{equation} \frac{dS}{dA} \simeq
\frac{min (\Delta S)}{min (\Delta A)} \simeq
\frac{1}{2L_p^2}\sqrt{\frac{\pi}{A}}\frac{1}{f_{disp}\left(\sqrt{\frac{4
 \pi}{A}}\right)}
 \label{minDaAO}
\end{equation}
and
\begin{equation}
T_{BH}\simeq \frac{1}{4 \pi }f_{disp}\left(\frac{E_p^2}{2M} \right) \label{BHT}
~.
\end{equation}

As illustrative examples of ``all-order MDRs" we consider
the following three cases:
\begin{eqnarray}
\cosh(E/E_p)-\cosh(m/E_p)-\frac{p^2}{2E_p^2}e^{E/E_p}&=&0,
\label{DSR1dd} \\
\frac{E^2}{(1-E/E_p)^2}-\frac{p^2}{(1-E/E_p)^2}-m^{2}&=&0,\label{DSR2dd}\\
\cosh(\sqrt{2} E/E_p)-\cosh(\sqrt{2} m/E_p)-\frac{p^2}{E_p^2}\cosh(\sqrt{2}
E/E_p)&=&0,\label{DSR3dd}
\end{eqnarray}
(\ref{DSR1dd}) has already been considered in the previous
literature~\cite{lukieAnnPhys,gacdsr,jurekdsr,gianluFranc},
particularly as a possible description of particle propagation
in $\kappa$-Minkowski noncommutative spacetime.
It provides an example in which the coefficient of the linear-in-$L_p$ term
is nonvanishing: $\alpha_1=-1/2$.
And it is noteworthy that according to (\ref{DSR1dd})
there is a maximum momentum for fundamental particles: from  (\ref{DSR1dd})
it follows that for $E\rightarrow \infty$ one has $p \rightarrow E_p$.

The case (\ref{DSR3dd}) has not been previously considered
in the literature. It provides for our purposes a valuable illustrative
example since, as in the case of (\ref{DSR1dd}),
it would lead to a maximum momentum ($p \rightarrow E_p$
 for $E\rightarrow \infty$) but, contrary to  the case of (\ref{DSR1dd}),
 it corresponds to $\alpha_1=0$ (whereas $\alpha_2=-5/18$).
 This is therefore an example with the maximum-momentum feature and
 such that one would expect the leading corrections to the entropy-area
 relation to be logarithmic.

The case of (\ref{DSR2dd}) has already been considered in the literature
for other reasons~\cite{leedsr}, and it provides us an opportunity
to illustrate some consequences of a scenario in which
both $\alpha_1$ and $\alpha_2$ vanish, but still there are some Planck-scale
modifications of the energy-momentum dispersion relation.
And it is noteworthy that (\ref{DSR2dd}) can be implemented in such a
way that~\cite{leedsr} the Planck scale provides the maximum value
of both momentum and energy.

For the cases with dispersion relations (\ref{DSR1dd}) or (\ref{DSR3dd}),
since $E\rightarrow \infty$ for $p \rightarrow E_p$, the formulas derived
above would lead to the conclusion that the black hole temperature
diverges at some finite (nonzero!) value of the black-hole
mass $M_{min}=E_p/2$. We would then assume that this $M_{min}$ is the
minimum allowed mass for a black hole, and that the standard
description of the evaporation process should not be applicable beyond
this small value of mass.

In cases in which one introduces both a maximum momentum and a maximum energy
while keeping the form of the dispersion relation largely
unaffected\footnote{Whenever
the mass $m$ can be ignored ({\it i.e.} for massless particles and high-energy
particles with finite mass) the dispersion relation (\ref{DSR2dd})
is indistinguishable from the standard special-relativistic one.},
as done in some applications of (\ref{DSR2dd}), one would expect
(since the energy has a maximum Planckian value, $E_{Max}=E_P$) that
the temperature should be bounded to be lower than the Planck scale,
 $T_{Max} \sim E_p$, and that the minimum allowed value of black-hole
 mass should be also Plankian, since it should be the value of mass
such that temperature reaches is maximum allowed value.

\section{GUP (with and without MDR) and black hole entropy}
In the previous section we focused on scenarios in which the
energy-momentum dispersion relation is modified but the position-momentum
uncertainty principle preserves the Heisenberg form.
But clearly the key ingredient of our analysis is the presence
of a correction term $\Delta$ in the particle-localization-limit
relation $E \ge 1/ \delta x + \Delta$.
As stressed in Section 2, both a MDR and a GUP can introduce
such a correction term in the particle-localization limit, and therefore,
as we want to discuss explicitly in this Section, also in presence of
a GUP one should expect corrections to the entropy-area black-hole formula
and to the formula that relates
the mass and the temperature of a black hole.

We start the section by considering
scenarios in which the position-momentum
uncertainty principle is Planck-scale modified, while the
energy-momentum dispersion relation preserves its special-relativistic form.
Then in Subsection 4.2 we comment on the more general case, in which one
might be dealing with both a MDR and a GUP.

\subsection{GUP and black hole entropy}
Let us start by noting here again for convenience the
particle-localization limit that one obtains assuming a GUP
of the form (\ref{GUP})
and a standard (special-relativistic) energy-momentum dispersion relation:
\begin{equation}
E \ge \frac{1}{\delta x} + \alpha \frac{L_p^2}{(\delta x)^3}
+ O\left(\frac{L_p^3}{(\delta x)^4}\right)\, .
\label{deformGUPbis}
\end{equation}

Following the same strategy of analysis adopted in the previous
section, one finds that the Bekenstein argument, when taking into
account this localization limit (\ref{deformGUPbis}),
leads to the conclusion
that the maximum increase of black-hole area
upon absorption of a particle of energy $E$ is given by
\begin{eqnarray}
\Delta A \ge 4 (\ln 2) \! \left[L_p^2
+\! \frac{\alpha L_p^4}{(\delta x)^2} \right] \! \simeq
 4 (\ln 2) \! \left[ L_p^2
+ \! \frac{ \alpha L_p^4}{(R_S)^2}  \right] \! \simeq  4 (\ln
2) \! \left[ L_p^2 + \! \frac{\alpha 4 \pi L_p^4}{A}
\right]\, . \nonumber
\end{eqnarray}

From this it follows that
the entropy-area relation should take the form
\begin{equation}
S \simeq \frac{A}{4 L_p^2} - \alpha  \pi \ln \frac{A}{L_p^2} ~, \label{finalGUP}
\end{equation}
and the formula relating the temperature and the mass of the black hole
should take the form
\begin{equation}
T_{BH}^{GUP}\simeq \frac{E_p^2}{8\pi M}\left(1+\alpha\frac{E_p^2}{8M^2} \right)~.
\end{equation}

\subsection{Combining MDR and GUP in the analysis of black-hole entropy}
We have argued that both a MDR and a GUP are
possible features of a quantum-gravity theory
that would affect black-hole termodynamics.
Actually, as the careful reader must have noticed, the line of analysis
we are advocating is composed of two steps. First we notice that
the ``particle-localization limit''
in its standard form, $E \ge 1/\delta x$, is derived on the basis of two
key assumptions, the validity of the Heisenberg
position-momentum uncertainty principle and
the validity of the special-relativistic energy-momentum
dispersion relation, and that by modifying the uncertainty principle
and/or the dispersion relation one gets a modified particle-localization
limit of the type $E \ge 1/\delta x + \Delta_{\delta x,L_p}$.
Then we observe that a key assumption of the Bekenstein argument for the
derivation of black-hole entropy is the validity
of the standard particle-localization limit $E \ge 1/\delta x$.
With a MDR and/or a GUP one gets a modified particle-localization limit,
which in turn leads to a modification of the black-hole area-entropy
relationship.

It is worth mentioning that the modifications induced by a MDR and a GUP may
(at least in part) cancel out at the level of the area-entropy equation.
In order to stress the importance of this possibility let us consider the
information presently available on the Loop Quantum Gravity approach:
(i) several Loop-Quantum-Gravity studies have argued in favour of a MDR
with nonvanishing $\alpha_1$ (leading Planck-scale correction to the
dispersion relation that goes linearly with $L_p$),
(ii) there is no mention
of a GUP in the Loop-Quantum-Gravity literature,
(iii) several Loop-Quantum-Gravity studies have
argued in favour of an entropy-area relationship
in which the leading correction, beyond the linear term,
is of log-area type.
According to the perspective on the derivation of black-hole entropy that
we are advocating one would find these three ingredients
to be logically incompatible:
if the MDR has nonvanishing $\alpha_1$ and the position-momentum uncertainty
principle is not Planck-scale modified then in the
entropy-area relationship the leading correction, beyond the linear term,
should have $\sqrt{area}$ dependence.
Does this mean that Loop Quantum Gravity is a logically
inconsistent framework?
Of course, it does not. It simply means that some of the relevant preliminary
results must be further investigated. It may well be that, as the loop-quantum-gravity
approach is understood more deeply, it turns out that the $\alpha_1$ coefficient in
the MDR vanishes. Or else we might discover that in Loop Quantum Gravity
the $\alpha_1$ coefficient in the MDR takes a nonzero value, but
there is a corresponding linear-in-$L_p$ term in the GUP with just the right
coefficient to give an overall vanishing coefficient to the $\sqrt{area}$ term
in the entropy-area relation.

Our perspective on the derivation of black-hole entropy provides a logical
link between different aspects of a quantum-gravity theory and may be
used most fruitfully when, as in the case of Loop Quantum Gravity,
the formalism is very rich and some of the results obtained within that formalism
are of preliminary nature.
Even before being able to derive more robust results we may uncover
that the presently-available preliminary results are not providing us
with a logically-consistent picture, and this in turn will give us additional
motivation for investigating more carefully those preliminary
results.

It is also worth mentioning that on the string-theory side our
perspective on the derivation of black-hole entropy provides no evidence
of a logical inconsistency among the results so far obtained in that framework.
The string-theory literature indicates that the entropy-area
relationship should involve a leading correction, beyond the linear term, of
log-area type, and provides strong evidence of a GUP of the type (\ref{GUP}),
while the results so far obtained do not indicate the need to modify
the dispersion relation in string theory. These three ingredients provide a
logically-consistent scenario within
our perspective on the derivation of black-hole entropy.
As shown above, with a GUP of the type (\ref{GUP}) and with an unmodified
(still special-relativistic) dispersion relation one is indeed led
to an entropy-area
relationship in which the leading correction, beyond the linear term, is of
log-area type.

\section{Implications for the Bekenstein entropy bound and Generalized Second Law}
It is natural at this point, after having shown that a MDR and
a GUP can affect the black-hole entropy-area and
mass-temperature relationships, to wonder whether other aspects
of black-hole thermodynamics are also affected, and whether the overall
picture preserves the elegance/appeal of the original scheme,
based on standard uncertainty principle and dispersion relation.
In this section we investigate the validity of the
Generalized Second Law (GSL) of thermodynamics and the implications
for the Bekenstein entropy bound.
In order to work within a definite scenario
we assume here a MDR (while we implicitly assume that the uncertainty
principle takes its standard form).

The GSL~\cite{Bekenstein:1974ax} asserts
that the second law of thermodynamics is still
valid in presence of collapsed matter. Given the entropy of the
black hole, as described by the area-entropy relation,
the GSL requires that the total entropy of a system
composed of a black hole and ordinary matter
never decreases.
This means that the following inequality holds
for all physical processes
\begin{equation}\label{GSL}
S_{BH}+S_{mat} \geq 0\, .
\end{equation}
It was observed~\cite{Bekenstein:1980jp} that
in principle (using the so-called ``Geroch process")
one could violate the GSL if objects of fixed size $R$ and
energy $E$ could have arbitrarily large entropy $S$.
This led Bekenstein to propose a ``entropy bound"
\begin{equation}
S_{mat}\leq2\pi E R
\end{equation}
for an arbitrary system of energy $E$ and effective radius $R$.
The fact that the GSL implies the Bekenstein bound and vice versa
has long been debated and is still actively debated.
However the Bekenstein bound turns out to hold for a variety
of systems in flat Minkowski space and can be derived as weak-gravity limit
of the popular ``Generalized Covariant Entropy Bound"~\cite{Bousso:2004kp}.

A remarkable feature of the Bekenstein bound is that, in spite of
being motivated by considerations rooted in the gravitational realm,
it does not involve the Planck scale (or equivalently Newton's constant).
The absence of the Planck scale is less puzzling in light of
the observation that the bound
can be derived even without advocating
gravity in any way: it is sufficient~\cite{calib2}
to analyze some implications
of the particle-localization limit $E\geq\frac{1}{\delta x}$.
This alternative derivation requires considering a
matter system with energy $E$,
in which self-gravitation effects can be neglected,
that occupies a region in flat spacetime with
radius $R$ smaller than the gravitational radius $R_{G} \equiv 2L_{p}^2E$.
The standard particle-localization limit, when generalized to this type
of systems, sets a minimum value for the energy of
a quantum in a region of spatial radius $R$
\begin{equation}
\epsilon(R)\geq\frac{1}{R}\, .
\end{equation}
The maximum number of quanta that we can have in the region is then given by
\begin{equation}
N_{max}\simeq\frac{E}{\epsilon(R)}=ER\, .
\end{equation}
If we consider the simple case of a system for which the
maximal number of microstates for $N$ particles is given
by $\Omega(N)=2^{N}$ then the entropy of the system $S=log\Omega(N)$
is bounded by the inequality
\begin{equation}
S_{mat} \leq \log 2 ER\, ,
\end{equation}
which is indeed consistent with the Bekenstein bound
(up to another "calibration factor" $\eta=\frac{2\pi}{log 2}$).

We briefly reviewed this derivation of the Bekenstein bound especially
in order to stress the role played by the particle-localization
limit $E\geq\frac{1}{\delta x}$. It is then obvious that the modifications
of the particle-localization limit induced by a MDR (and/or a GUP)
would affect the Bekenstein bound.
As shown earlier, within our parametrization
of the MDR\footnote{For an analogous
modification of the Bekenstein bound coming from
the GUP see Ref.~\cite{calib2}.},
one obtains a particle localization limit of the form
\begin{equation}
\epsilon(R)\geq \frac{1}{R} \left(1-\alpha_{1}\frac{L_p}{R}
-\left(\frac{3}{2} \alpha_2-\frac{11}{8}  \alpha_1^2\right)\frac{L_p^2}{R^2}+O\left(\frac{L_p^3}{R^3}\right)\right)
\end{equation}
which gives
\begin{equation}
S_{mat} \leq 2\pi ER \left(1+\alpha_{1}\frac{L_p}{R}+\left(\frac{3}{2} \alpha_2-\frac{11}{8}  \alpha_1^2\right)\frac{L_p^2}{R^2}
+O\left(\frac{L_p^3}{R^3}\right)\right)\,.
\label{joc99}
\end{equation}

This MDR-modified Bekenstein bound fits very naturally with our
corresponding formula, (\ref{entropy1}),
for the entropy-area relation; in fact, the two results combine to
provide us with a picture which is still consistent with the GSL.
According to (\ref{joc99}) when a matter system of energy $E$
falls into the black hole, this corresponds to a negative change of entropy
which has absolute value not greater than
\begin{equation}
max(|\Delta S_{mat}|)\simeq 2\pi ER \left(1+\alpha_{1}\frac{L_p}{R}
+\left(\frac{3}{2} \alpha_2-\frac{11}{8}  \alpha_1^2\right)\frac{L_p^2}{R^2}+O\left(\frac{L_p^3}{R^3}\right)\right)
\end{equation}
and correspondingly, according to  (\ref{entropy1}),
the black hole entropy increases at least by
 \begin{equation}
min(\Delta S_{BH})\simeq 2\pi ER \left(1+\alpha_{1}\frac{L_p}{R}
+\left(\frac{3}{2} \alpha_2-\frac{11}{8}  \alpha_1^2\right)\frac{L_p^2}{R^2}+O\left(\frac{L_p^3}{R^3}\right)\right)
\end{equation}
Thus the MDR-induced corrections to $S_{BH}$ and $S_{mat}$ cancel exactly
at the level of the inequality relevant for the GSL.
The GSL stills holds, even in presence of a modified particle-localization
limit.

\section{Corrections to black-body radiation spectrum}
In preparation for some observations on black-hole evaporation,
to which we devote Section 7, we now want to investigate the implications
of a MDR and/or a GUP for the
black-body radiation spectrum.

\subsection{MDR and black-body spectrum in leading order}
Let us start by considering photons in a cubical box with edges of length $L$
(and volume $V=L^3$).
The wavelengths of the photons are subject to
the boundary condition $\frac{1}{\lambda}=\frac{n}{2L}$, where $n$ is a
positive integer.
This condition implies, assuming that the de Broglie relation
is left unchanged, that the photons have (space-)momenta
that take values $p=\frac{n}{2L}$. Thus
momentum space is divided into cells of
volume $V_p=\left(\frac{1}{2L}\right)^3=\frac{1}{8V}$.
From this it follows that the number of modes with momentum
in the interval $[p,p+dp]$ is given by
\begin{equation}
g(p) dp =8\pi V p^2 dp\, .
\label{joc87}
\end{equation}
Assuming a MDR of the type parametrized in (\ref{disprelONE}) one then
finds that ($m=0$ for photons)
\begin{equation}
p\simeq E \left(1+\frac{\alpha_1}{2} L_p E+\left(\frac{\alpha_2}{2}-\frac{\alpha_1^2}{8}\right) L_p^2 E^2\right)
\end{equation}
and
\begin{equation}
dp \simeq  \left(1+ \alpha_1 L_p E+\left(\frac{3}{2}\alpha_2-\frac{3}{8}\alpha_1^2\right) L_p^2 E^2\right) dE
\end{equation}
Using this in (\ref{joc87}) one obtains
\begin{equation}
g(E) dE = 8\pi V \left(1+2 \alpha_1 L_p E
+5\left(\frac{1}{2}\alpha_2+\frac{1}{8}\alpha_1^2\right) L_p^2
E^2\right) E^2 dE\,
\end{equation}
which in terms of the frequency $\nu$ takes the form
\begin{equation}
g(\nu) d\nu = 8 \pi V \left(1+2 \alpha_1 L_p \nu
+5\left(\frac{1}{2}\alpha_2+\frac{1}{8}\alpha_1^2\right) L_p^2
\nu^2\right) \nu^2 d\nu\, .
\label{joc86}
\end{equation}

In order to obtain the MDR-modified
energy density of a black body at temperature $T$
we must now use (\ref{joc86}) and rely on the statistical
arguments which show that in a
system of bosons at temperature $T$ the average energy per oscillator
is given by
\begin{equation}
\bar{E}=\frac{\nu}{e^{\frac{\nu}{T}}-1}
~.
\label{aveE}
\end{equation}
Thus the energy density at a given temperature $T$,
for the frequency interval $[\nu, \nu+d\nu]$, is
\begin{equation}
u_\nu (T) d\nu= 8\pi \left(1+2 \alpha_1 L_p \nu
+ 5\left(\frac{1}{2}\alpha_2+\frac{1}{8}\alpha_1^2\right) L_p^2
E^2\right) \frac{\nu^3 d\nu}{e^{\frac{\nu}{T}}-1}\, . \label{densitynu1}
\end{equation}
and integrating this formula
we get the MDR-modified energy density of a black body at temperature $T$
\begin{equation}
u (T)= \frac{8\pi^5}{15} T^4 +384 \pi \zeta(5) \alpha_1 L_p T^5
+5\left(\frac{1}{2}\alpha_2+\frac{1}{8}\alpha_1^2\right)
 \frac{160\pi^7}{63}  L_p^2 T^6 \label{enu1}
\end{equation}
The MDR introduces corrections
of the type $T^{4+n}/E_P^n$ to the Stefan-Boltzmann law.
Moreover, the maximum value of the integrand in (\ref{densitynu1}),
as a function of $\nu$, is clearly also shifted: the MDR also introduces
a modification of Wien's law.
Of course, using the low-energy expansion (\ref{disprelONE})
of the dispersion relation we only get a reliable picture at temperatures
safely below the Planck scale, but the presence of correction terms
of the type $T^{4+n}/E_P^n$  clearly suggests that the MDR-modified description
leads to departures from the Stefan-Boltzmann law that can become very
significant as the temperature approaches the Planck scale.
We intend to show this explicitly by considering an example
of all-order MDR formula.

\subsection{Some all-order results for MDR modifications of black-body spectrum}
Let us therefore derive once again the modified Stefan-Boltzmann law,
now assuming, as illustrative example of an all-order MDR formula,
the validity of the dispersion relation (\ref{DSR3dd}).
Clearly the number of modes in momentum space is still given by
\begin{equation}
g(p)dp=8\pi V p^2 dp\, ,
\end{equation}
but now
\begin{equation}\label{disprelall}
p^2=E_p^2\left(1-\frac{1}{\cosh(\sqrt{2}E/E_p)} \right)
\end{equation}
and this implies that the number of modes for given energy is given by
\begin{equation}
g(E)dE=16 \pi V E_p^2 \sinh^2\left(\frac{E/E_p}{\sqrt{2}} \right)
\cosh\left(\frac{E/E_p}{\sqrt{2}}\right)
 \frac{1}{\cosh^{5/2}\left(\sqrt{2}E/E_p \right)} dE\,
\end{equation}
{\it i.e.} the number of modes for given frequency is
\begin{equation}
g(\nu)d\nu=16 \pi V E_p^2 \sinh^2\left(\frac{\nu/E_p}{\sqrt{2}} \right)
\cosh\left(\frac{\nu/E_p}{\sqrt{2}}\right)\frac{1}{\cosh^{5/2}
 \left(\sqrt{2}\nu/E_p \right)}  d\nu\
 ~.
 \label{joc72}
\end{equation}

Then the modified Stefan-Boltzmann law is given, in integral form,
by
\begin{equation}
u (T) = \frac{1}{V} \int_0^\infty \frac{g(\nu)}{e^{\frac{\nu}{T}}-1}\nu d\nu
~,
 \label{densitynuall}
\end{equation}
where $g(\nu)$ is the one of (\ref{joc72}).

It is useful to consider some limiting forms of
the integration in (\ref{densitynuall}).
Clearly, since (\ref{DSR3dd}) is consistent with  (\ref{disprelONE})
for $\alpha_1=0$
and $\alpha_2=-5/18$, in the limit $T/E_p\ll1$ the integration (\ref{densitynuall})
gives a result that reproduces (\ref{enu1}) for $\alpha_1=0$
and $\alpha_2=-5/18$.
But, now that we are dealing with an all-order formula,
besides considering the case $T/E_p\ll1$ we can
also investigate the opposite limit $T/E_p\gg1$, finding
\begin{equation}\label{TEgg1}
u (T) = 16 \pi E_p^4\left\{\frac{T}{E_p}C_1
-\frac{1}{2}C_2-\frac{E_p}{T}C_3+O(E_p^2/T^2)\right\}
\end{equation}
where
\begin{eqnarray}\label{Cs}
C_1&=& \int_0^\infty \sinh^2(x/\sqrt{2})
\frac{\cosh(x/\sqrt{2})}{\cosh^{5/2}(\sqrt{2}x)}dx=\frac{1}{6}, \\
C_2&=& \int_0^\infty x \sinh^2(x/\sqrt{2})
\frac{\cosh(x/\sqrt{2})}{\cosh^{5/2}(\sqrt{2}x)}dx \simeq 0.22,\\
C_3&=& \int_0^\infty x^2 \sinh^2(x/\sqrt{2})
 \frac{\cosh(x/\sqrt{2})}{\cosh^{5/2}(\sqrt{2}x)}dx \simeq 0.41,
\end{eqnarray}
This means that the MDR (\ref{DSR3dd}) leads to a modification of the
Stefan-Boltzmann law which at the Planck scale is very significant:
for $T \gg E_p$ one finds that $u$ depends linearly on $T$, rather than
with the fourth power.

It is of particular interest to establish what is the relationship between
the ``characteristic frequency'' (and characteristic wavelength)
of the black-body spectrum and temperature.
In the standard description of a black body the
characteristic frequency grows linearly with the temperature.
In order to verify whether this is still the case in our MDR-modified
scenario we can
take the derivative of $u_\nu (T)$ with respect to $\nu$, so that we can
identify the value of frequency
for which the energy density (and the radiated flux) reaches a maximum.
This leads to the following equation that must
be satisfied by the characteristic frequency $\bar{\nu}$:
\begin{equation}
\left(e^{\frac{\bar{\nu}}{T}}-1\right)\left(g(\bar{\nu})+g'(\bar{\nu})\bar{\nu}\right)
-\frac{e^{\frac{\bar{\nu}}{T}}}{T}g(\bar{\nu})\bar{\nu}=0 ~.
\end{equation}
For $T \ll E_p$ of course this reproduces the type of small
modification of Wien's law, which we already noticed in the
previous section.
The fact that we are now considering a scenario with
a given all-order MDR formula
allows us to examine the dependence of the
characteristic frequency on temperature
even when the temperature reaches and eventually exceeds the Planck scale.
And we find that the for $T/E_p\gg1$ the characteristic frequency becomes
essentially independent of temperature.
No matter how high the temperature goes the characteristic frequency
never exceeds the following finite value:
\begin{equation}
\bar{\nu}\simeq E_p \frac{\cosh^{-1}[(1+\sqrt{41})/4]}{\sqrt{2}} \simeq 0.87 E_p
\label{barnu}
\end{equation}
So basically at low temperatures any increase of temperature causes a
corresponding increase in characteristic frequency of the black-body spectrum,
but gradually a saturation mechanism takes over and even in the
infinite-temperature limit the characteristic frequency is still finite,
and given by the Planck scale (up to a coefficient of order 1).
This occurs with the dispersion relation (\ref{DSR3dd}), {\it i.e.}
in a scenario with a minimum value of wavelength but no maximum value
of frequency.
An analogous result for the case of the dispersion relation (\ref{DSR2dd}),
which leads to both a minimum value of wavelength and a maximum value of
frequency, would have not been surprising: if the framework introduces
from the beginning a maximum Planckian value of frequency, then of course
also the characteristic frequency of black-body radiation would
be ``subPlanckian".
But in analyzing the case of (\ref{DSR3dd}) we found that the presence of
a minimum wavelength at the fundamental level is sufficient for the emergence
of  a maximum Planckian value of the characteristic frequency of black-body
radiation, as shown explicitly by Eq.~(\ref{barnu}).


\subsection{Black-body spectrum with GUP}
In the previous two subsections the key point was that a MDR leads to
a modified formula for the density of modes in a given (infinitesimal)
frequency interval, $g(\nu) d\nu$.
If instead we now assume that the dispersion relation takes its standard
special-relativistic form, but there is a GUP, it is not {\it a priori}
obvious that the black-body spectrum is affected.
One does indeed obtain a modified black-body spectrum if
it is assumed that the GUP should also be reflected in a
corresponding modification of the de Broglie relation,
\begin{equation}
\lambda\simeq \frac{1}{p}\left(1+\alpha L_p^2 p^2\right)
\end{equation}
and
\begin{equation}
E\simeq \nu \left(1+\alpha L_p^2 \nu^2\right)\, . \label{modplanck}
\end{equation}

For oscillators in a box the number of modes in an
infinitesimal frequency interval would still be described by
the standard formula
\begin{equation}
g(\nu) d\nu = 8 \pi V \nu^2 d\nu\, ,
\end{equation}
but, as a result of (\ref{modplanck}), the average energy per oscillator
would be given by
\begin{equation}\label{aveEmod}
\bar{E}=\frac{\nu}{e^{\frac{\nu}{T}}-1} \left(1+\alpha L_p^2
\nu^2\left(1-\frac{\frac{\nu}{T}}{1-e^{-\frac{\nu}{T}}}\right)\right)\, .
\end{equation}

Combining (\ref{modplanck}) and (\ref{aveEmod}) one finds
\begin{equation}
u_\nu (T) d\nu= 8\pi \left(1+\alpha L_p^2\nu^2 \left(1-\frac{\frac{\nu}{T}}{1
-e^{-\frac{\nu}{T}}}\right)\right)
\frac{\nu^3 d\nu}{e^{\frac{\nu}{T}}-1}\, . \label{densitynu1bis}
\end{equation}
and the modified Stefan-Boltzmann law takes the form
\begin{equation}
u (T)=  \frac{8\pi^5}{15} T^4 + \frac{8\pi^6}{9}\alpha L_p^2 T^6\,\, .
\end{equation}
The $L_p^2 T^6$ correction term is just one of the $L_p^n T^{4+n}$
correction terms on which we already commented in the context of the
MDR modifications
of black-body radiation.

\section{Black hole evaporation}
In this section we use some of the results obtained in the previous
sections in a description of the black-hole evaporation process.
The key
ingredients are the relation between the black-hole temperature and mass and
the relation between the black-hole temperature
and the energy density emitted by the black hole.
We neglect possible non-thermal corrections
due to back-reaction effects (see, {\it e.g.}, the recent
studies in Ref.~\cite{michLEAK,vageLEAK} and references therein),
and we therefore treat the radiation emitted by the black-hole
as black-body radiation.

\subsection{MDR and Black hole evaporation}
At temperature $T$ the intensity $I$ of the radiation emitted by
a black hole of area $A$ is given by
\begin{equation}
I(T)=A \, u(T)
~.
\end{equation}

Using energy conservation one can write
\begin{equation}\label{eva1}
\frac{dM}{dt}=-A \, u,
\end{equation}
and assuming a MDR of the type $E=f_{disp}(p)$, in light of
our result (\ref{BHT}), one finds
\begin{equation}
\frac{dM}{dt}=-
16 \pi  \frac{M^2}{E_p^4}
u\left(\frac{1}{4\pi}f_{disp}\left(\frac{E_p^2}{2M}\right) \right)
\end{equation}

When $M\gg E_p$ (so that a power-series expansion of $f_{disp}(E_p^2/2M)$
is meaningful) this takes the form
\begin{eqnarray}\label{bhelo}
\frac{dM}{dt}=&=& - k_0\frac{E_p^8}{M^2} - k_1 \alpha_1 \frac{E_p^9}{M^3}
     - (k_{21}\alpha_1^2 + k_{22}\alpha_2)
    \frac{E_p^{10}}{M^4}+O(E_p^5/M^5)
\end{eqnarray}
where $k_0=\frac{\pi^2}{480}$, $k_1=k_0\frac{90\zeta(5)-\pi^5}{\pi^5}$,
 $k_{21}=k_0\frac{502 \pi^5-75600\zeta(5)}{672 \pi^5}$
 and $k_{22}=-k_0\frac{211}{672 \pi^5}$

This power-series analysis allows to conclude that a MDR can affect
the speed of evaporation of a black hole. For example,
in the case of the dispersion relation (\ref{DSR1dd})
the evaporation process is retarded with respect to the standard case,
whereas in the case of (\ref{DSR3dd}) the evaporation process is
accelerated.

With a given all-order MDR formula one can obtain of course even more
detailed information than available using the power-series expansion.
In particular, let us look at the case of the dispersion relation (\ref{DSR3dd})
and analyze the stage of the evaporation process when the mass of the
black hole is of the order of the Planck scale.
For $M\sim E_P$ we can approximate the MDR (\ref{DSR3dd}) as follows
\begin{equation}\label{eexp}
    E \simeq \frac{E_p}{\sqrt{2}} \ln
    \left(\frac{2}{1-(p/E_p)^2}\right)
\end{equation}
and then one finds
\begin{equation}
\frac{dM}{dt} \simeq - (16 \pi)^2 M^2\left\{\frac{C_1}{4 \pi
\sqrt{2}}\ln\left(\frac{2}{1-(\frac{E_p}{2M})^2}\right)-\frac{1}{2}C_2 \right\}
 \label{bheao}
 ~.
\end{equation}
This shows that, in the case of the MDR (\ref{DSR3dd}),
the energy flux emitted by the black hole would formally
diverge as the  black-hole mass approaches $E_p/2$.
This is mainly a consequence of the fact that
the black-hole temperature diverges when $M\rightarrow E_p/2$.
In the standard description of black-hole evaporation
these divergences occur as $M \rightarrow 0$.

\subsection{GUP and Black hole evaporation}
The observations reported in the previous subsection for the
case of a MDR (with unmodified energy-momentum uncertainty relation)
can be easily adapted to the complementary situation with
a GUP and a standard (unmodified) dispersion relation.
One must however assume, as already stressed in Subsection 6.3,
that the GUP is reflected in a
corresponding modification of the de Broglie relation
($\lambda\simeq (1+\alpha L_p^2 p^2)/p$). In this hypothesis one
easily finds that
the black hole should lose its mass at a rate given by
\begin{equation}
\frac{dM}{dt}=-A\, u\left(\frac{E_p^2}{2M}\right)=
-16 \pi \frac{8\pi^5}{15} \left(T\left(\frac{E_p^2}{2M}\right)\right)^4
+ \frac{8\pi^6}{9}\alpha
L_p^2 \left(T\left(\frac{E_p^2}{2M}\right)\right)^6\,.
\end{equation}
Expanding for $M/E_p\gg1$ we obtain
\begin{equation}
\frac{dM}{dt} \simeq
-16 \pi \frac{E_p^4}{M^2} \left(\tilde{k}_0
+\alpha\tilde{k}_1\frac{E_p^2}{M^2} \right)
~,
\end{equation}
with $\tilde{k}_0=\frac{\pi}{7680}$
and $\tilde{k}_1=\frac{1}{294912}+\frac{\pi}{15360}$.

Clearly the modifications to the black hole
evaporation formula obtained in the GUP scenario
are qualitatively the same as in the MDR scenario
with $\alpha_1=0$.

\section{A possible dependence on the speed law for photons}
Throughout our analysis
we have implicitly assumed that the law $v_\gamma = 1$ describing the
speed of photons is not affected by the MDR and/or the GUP.
The possibility of modifications of the speed law for photons
has been however considered rather extensively, particularly in
the MDR literature.
While several authors have argued that the law $v_\gamma = 1$ should
not be modified even in presence of
an MDR~(see, {\it e.g.},
Refs.~\cite{Kosinski:2002gu,Mignemi:2003ab,Daszkiewicz:2003yr}
and references therein),
one also finds support in the literature for the
proposal (see, {\it e.g.}, Ref.~\cite{gianluFranc} and references therein)
of the law $v_\gamma = [dE/dp]_{m=0}=[df_{disp}(p)/dp]_{m=0}$
and the proposal
(see, {\it e.g.}, Ref.~\cite{leedsr} and references therein)
of the law $v_\gamma = p/E$.

For our analysis  a key point is that if, instead of $v_\gamma = 1$,
one took $v_\gamma = [dE/dp]_{m=0}$
or $v_\gamma = p/E$ then the speed of photons would acquire an
energy dependence which should be taken into account in some
aspects of our derivations.
We postpone to future studies this more general analysis, but in order to explore
the type of modifications which could be induced by
such an energy dependence of the speed
of photons we do intend to consider here the description of black-body radiation
with the dispersion relation (\ref{DSR3dd}), assuming
that the speed of photons is governed by either $v_\gamma = [dE/dp]_{m=0}$
or  $v_\gamma = p/E$.

We focus on the emitted ``flux density"
\begin{equation}
I_{\nu}=A\,u_{\nu}\,v_\gamma(\nu)
\end{equation}
where $A$ is the area of the radiating surface
and $u_{\nu}$ is the energy density at a
given frequency.

Taking $v_\gamma = p/E$, from (\ref{DSR3dd}) it follows that
\begin{equation}
v_\gamma(\nu)=\frac{p}{E}=\frac{E_p}{E}\sqrt{1
-\frac{1}{\cosh\left(\frac{\sqrt{2}E}{E_p}\right)}}
~.
\end{equation}
From this it would then follow that
the energy flux density is given by
\begin{equation}
I_\nu(T)=4\pi A \sqrt{2}E_p^3
\frac{1}{e^{\nu/T}-1}\frac{\sinh(\sqrt{2}E/E_p)}{\cosh^3(\sqrt{2}
E/E_p)}\left[\cosh(\sqrt{2}E/E_p)-1\right]
~. \label{new77}
\end{equation}
This suggests that,
although there are some small quantitative differences,
the qualitative features of black-body radiation
with the dispersion relation (\ref{DSR3dd})
are largely independent of the choice between $v_\gamma=1$
and $v_\gamma = p/E$.
In particular, from (\ref{new77}) with one finds that
the typical frequency of the photons contributing
to the energy flux saturates at
\begin{equation}
\bar{\nu}\simeq 0.76 E_p ,
\end{equation}
which is not much different from the typical frequency found for
the case $v_\gamma = 1$.
The analysis of the total emitted
energy ($\int_0^\infty I_\nu(T)d\nu $) also leads to rather small differences
between the choices $v_\gamma=1$
and $v_\gamma = p/E$. In particular from (\ref{new77}) one finds
\begin{equation}
I/A=\frac{8}{15}\pi^5 T^4\left\{1+
C_1\left(\frac{T}{E_p}\right)^2+C_2\left(\frac{T}{E_p}\right)^4
+O\left(\frac{T}{E_p}\right)^6\right\},
\end{equation}
in the limit $T/E_p\ll1$, and
\begin{equation}
I/A=E_p^4\left\{\widetilde{C}_1\frac{T}{E_p}+\widetilde{C}_2
+\widetilde{C}_3\frac{E_p}{T}
+O\left(\frac{E_p}{T}\right)^{2}\right\},
\end{equation}
in the limit $T/E_p\gg1$,
where $C_1=-\frac{100\pi^2}{21}$, $C_2=\frac{164\pi^4}{5}$
and ${\widetilde{C}_1}=5.57,{\widetilde{C}_1}=-\pi$
and ${\widetilde{C}_3}=0.79$ .

If instead one adopts the law $v_\gamma = [dE/dp]_{m=0}$, still
assuming (\ref{DSR3dd}),
one obtains
\begin{equation}
v_\gamma(\nu)=\frac{dE}{dp}=\frac{\cosh^{\frac{3}{2}} {\frac{\sqrt{2}
E}{E_p}}}{\cosh {\frac{E}{\sqrt{2}E_p}}}
\end{equation}
and then the flux density takes the form
\begin{equation}
I_{\nu}=16\pi A E_{p}^2\,\nu\frac{\sinh^2{\frac{\nu}{\sqrt{2}
E_p}}}{\left(e^{\frac{\nu}{T}}-1\right)\cosh{\frac{\sqrt{2}\nu}{E_p}}}\, .
~\label{last}
\end{equation}
From this one easily verifies that the effects
induced by the Planck-scale deformation in the case $v_\gamma = [dE/dp]_{m=0}$
are essentially of the same type encountered in the cases $v_\gamma = 1$
and $v_\gamma = p/E$, but the quantitative differences between the
case $v_\gamma = [dE/dp]_{m=0}$ and the other two cases are more significant
then the ones between the cases $v_\gamma = 1$
and $v_\gamma = p/E$.
As mentioned, in absence of the Planck-scale effects
the typical frequency of the photons contributing
to the energy flux grows linearly with the temperature, while in the
cases in which the Planck-scale effects of (\ref{DSR3dd}) are introduced
with $v_\gamma = 1$ or $v_\gamma = p/E$ the typical frequency
saturates at a Planckian value. If the same Planck-scale effects are introduced
with $v_\gamma = [dE/dp]_{m=0}$, as implicitly codified in (\ref{last}),
one finds that the growth of the typical frequency with temperature also slows down
significantly at high temperatures but it does not completely saturate:
at high temperatures the typical frequency
grows logarithmically with the temperature.

In summary the choice of the speed law does not appear to affect the core
features of the analysis, but it appears that it could in some cases
introduce some significant quantitative differences.

\section{Comparison with previous analyses}
To our knowledge, the one we reported here, in spite of its preliminary nature,
 is at this point
 the most composite effort of exploration of the implications of a MDR and/or
 a GUP in black-hole thermodynamics.
 But parts of the overall picture we attempted to provide had been investigated
 previously, and it seems appropriate to comment briefly on this previous
 related studies.

Closest in spirit to our perspective are the studies of the implications
of the GUP for black-hole thermodynamics reported in Refs.~\cite{Adler:2001vs}
and~\cite{calib2}.
Whereas for us (\ref{GUP}) is to be handled prudently, as it could possibly
be only an approximate form of a more complicated all-order-in-$L_p$ formula,
in Refs.~\cite{Adler:2001vs,calib2} the
formula (\ref{GUP}) is taken as the exact form
of the GUP, thereby leading to a corresponding form of the entropy-area
relation.
Perhaps more importantly Refs.~\cite{Adler:2001vs,calib2} assume that the
GUP would not affect the black-body spectrum and in
particular a standard expression for Stefan's law is used
even in Planckian regimes.
There was no investigation of MDRs in Refs.~\cite{Adler:2001vs,calib2}.

An attempt to describe Hawking radiation
in presence of a MDR was reported
in Ref.~\cite{Blaut:2001fy}.  There the problem is approached from
the field-theoretic perspective, considering
possible modification of the field equations coming from the
MDR.  No explicit formula for the corrections
to the Hawking spectrum and to the entropy-area relation was obtained
in Ref.~\cite{Blaut:2001fy}.

Ref.~\cite{Chang:2001bm} investigates
how a general form of the GUP could modify the volume element of phase
space, and therefore the black-body-radiation formula,
using the Hamiltonian formulation in terms of
Poisson brackets.

In Ref.~\cite{Yepez:2004sa} an analysis of black-body radiation
is carried out in presence of a MDR of the type emerging
from a proposed ``semiclassical limit" of Loop Quantum Gravity,
which is analogous to the ``leading order'' MDR (\ref{disprelONE})
we studied in some parts of this paper.  The results reported
there are consistent with the power-series formulas for Stefan's
and Wien's law which we derived. The features we exposed in considering some
illustrative examples of all-order MDRs,
were not discussed in Ref.~\cite{Yepez:2004sa}.
Also the entropy-area relation and the aspects
of black-hole evaporation which we considered here were not part of
the analysis reported in Ref.~\cite{Yepez:2004sa}, and Ref.~\cite{Yepez:2004sa}
did not consider the possibility of a GUP.

Ref.~\cite{am1} is closest in spirit to the part of our analysis where
we focused on the black-body radiation spectrum, as affected by a MDR.
Although the formal setup differs in several points,
the results are roughly consistent with ours, including
the possibility of ``saturation'' of the characteristic frequency
at $T \gg E_p$.
There was however no investigation of the entropy-area relation and
the Generalized Second law in Ref.~\cite{am1}, and Ref.~\cite{am1} also
did not consider the possibility of a GUP.

\section{Outlook}
The technical difficulties that are encountered in most approaches to
the quantum-gravity problem usually only allow one to grasp a few
disconnected aspects of the physical picture that the theories could provide.
And in some approaches even the few ``physical" results that are obtained,
are only derived within approximation schemes whose reliability is not
fully established.
We have argued that in this situation it might be particularly valuable to
establish a few logical links connecting some apparently unrelated
aspects of the physical picture. And we showed that such a link can
be found between some aspects of quantum-gravity research which have
attracted strong interest in recent times, a link providing a connection
between results on modified energy-momentum dispersion relations
and/or modified position-momentum uncertainty principles and results
on the thermodynamics of black holes.
We have provided a description of log corrections to the entropy-area law
for black holes that is based on the availability of a MDR and/or a GUP.

In exploring other aspects of black-hole thermodynamics as affected by
MDRs and GUPs we stumbled upon a few noteworthy points. We found that
the Generalized
Second Law of thermodynamics might be robust enough to survive the
introduction of these Planck-scale effects.
We found that a MDR introducing a minimum value for wavelengths
(even when no maximum value for frequencies is introduced)
could lead to a description of black-body radiation in which
the characteristic frequency of the radiation
never exceeds a finite Planckian value (described in Eq.~(\ref{barnu})).
This in turn also affects black-hole evaporation in such a way that
the temperature diverges already when the mass of the black hole decreases
to a Planck-scale value (instead of diverging only in the zero-mass limit
as usually assumed).

A key test for our line of analysis will come from future improved
analyses within the loop-quantum-gravity approach. According to
the perspective we adopted some preliminary results on the emergence
of modifications of the dispersion relation that depend linearly
on the Planck length (at low energies) would be incompatible with the
loop-quantum-gravity results on log corrections to the entropy-area relation
for black holes.
We predict that improved analyses of the loop-quantum-gravity approach
should lead to the emergence of a picture that is instead compatible
with the conceptual link we are proposing.

As stressed in Section~8 one aspect of our analysis in which we took
a rather conservative attitude (in comparison with the possibilities
considered in the literature) is the one concerning the description
of the speed of photons, which we assumed to be still frequency
independent. We do not expect major obstacles for a generalization
of our analysis with the inclusion of the possibility of a frequency-dependent
speed of photons, and the preliminary investigation reported in Section~8
suggests that some of the core features that emerged from our analysis
are only moderately affected by the choice of law for the  speed of photons.

\section*{Acknowledgments}
The work of
M.~A.~was supported by a Fellowship from The Graduate School
of The University of North Carolina. M.~A.~also
thanks the Department of Physics of the University of Rome for hospitality.
Y.~L.~is partly supported by NSFC (No.10205002,10405027) and
SRF for ROCS,SEM.

\end{document}